# Molecular beam epitaxy of wafer-scale O-band InAs/InGaAs quantum dots on GaAs for quantum photonics

Pavel S. Avdienko,[1, a)] Lukas Hanschke,[2] Quirin Buchinger,[3, 4] Nikolai Bart,[5] Hubert Riedl,[1] Bianca Scaparra,[2] Yu Xia,[6] Ziria Herdegen,[6] Knut Müller-Caspary,[6] Gregor Koblmüller,[1, 7] Tobias Huber-Loyola,[3, 4, 8] Arne Ludwig,[5] Andreas Pfenning,[3, 4] Sven Höfling,[3] Kai Müller,[2] and Jonathan J. Finley[1, b)]

[1)]*Walter Schottky Institut, Physik Department, School of Natural Sciences, Technical University of Munich, Am Coulombwall 4, 85748 Garching b. München, Germany*
[2)]*TUM School of Computation, Information and Technology, Technical University of Munich, 80333 Munich, Germany*
[3)]*Julius-Maximilians-Universität Würzburg, Physikalisches Institut, Lehrstuhl für Technische Physik, Am Hubland, 97074 Würzburg, Germany*
[4)]*Saqura Technologies GmbH, Paradiesstr. 15, 97855 Homburg am Main, Germany*
[5)]*Ruhr University Bochum, Faculty of Physics and Astronomy, Experimental Physics VI, 44801 Bochum, Germany*
[6)]*Department of Chemistry and Center for NanoScience, Ludwig-Maximilians-University Munich, 81377 Munich, Germany*
[7)]*Institute of Physics and Astronomy, Technical University Berlin, Hardenbergstrasse 36, 10623 Berlin*
[8)]*Karlsruhe Institute of Technology, Institute of Photonics and Quantum Electronics IQST, Engesserstr. 5, 76131 Karlsruhe, Germany*

(Dated: 11 February 2026)

We report a scalable molecular beam epitaxy strategy to achieve a low density of O-band electrically tunable InAs/InGaAs quantum dots (QDs) on GaAs(001) substrates. Our approach is based on a gradient deposition of InAs in the sub-ML regime and subsequent capping with an $In_{0.29}Ga_{0.71}As$ strain-reducing layer to redshift the emission wavelength. For different growth conditions, we investigate the optical properties of the dots using photoluminescence mapping and correlate with structural properties determined by scanning transmission electron microscopy. Using a surface roughness modulation technique and synchronizing InAs sub-monolayer deposition cycles with substrate rotation, we control the dot density and position low-density regions ($< 10^8$ cm$^{-2}$) on the substrate. Hyperspectral imaging is used to map the spatial and spectral characteristics of many individual dots in the low-density region, confirming that our approach is universally applicable to conventional MBE growth on (001) surfaces. Finally, we tune the QD emission wavelength within the O-band using electric fields and demonstrate single-photon emission with $g^{(2)}(0) = 0.020 \pm 0.014$.

## I. INTRODUCTION

Epitaxial semiconductor quantum dots (QDs) can be employed as efficient single-photon sources (SPSs) required for photonic quantum technologies, such as quantum communication networks and quantum computing[1–5]. The majority of studies have focused on InAs/GaAs quantum dots emitting at wavelengths below 1 µm, whereas fiber-based quantum technologies compatible with existing telecommunication infrastructure require SPSs with high photon indistinguishability and extraction efficiency operating in the telecoms O-band (~1.3 µm) or C-band (~1.55 µm), respectively[1,6]. Efficient SPSs call for the incorporation of QDs into optical cavities for practical applications with high-rate single-photon emission. The well-established technology and fabrication scalability of GaAs-based photonic heterostructures, including electrical control, make them particularly advantageous for photonic quantum technologies[3,5,7–9]. However, simultaneously achieving emission in the telecom bands and a low density of InAs QDs grown on GaAs substrates remains challenging due to the difficulty of growing sufficiently large and In-rich dots[7,10,11].

The low-temperature emission wavelength of self-assembled InAs/GaAs QDs grown in Stranski-Krastanow (SK) growth mode and the growth temperature range 480-520°C does not usually exceed 1.2 µm[7,12,13]. Several approaches have been followed to redshift the emission of InAs QDs grown on GaAs substrates into the telecommunication spectral windows. These include, (i) the overgrowth of InAs QDs grown on GaAs by an In(Al,Ga)As(N) layer (the so-called strain-reducing layer (SRL))[5,7,10,14–16], (ii) embedding InAs QDs grown in symmetric or asymmetric In(Al,Ga)As quantum wells[17,18] and (iii) growth of InAs QDs on metamorphic In(Al,Ga)As buffer layers[19–21]. Each of these approaches is based on the possibility of engineering band profiles leading to a redshift of the QD emission through the reduction of the compressive strain of InAs/Ga(In)As QDs and/or band-offset[7,10,17–21]. It has been demonstrated that these approaches can be used for the development of nanophotonic devices with high single-photon emission efficiency[5,22,23], integration with fiber-optic infrastructure[9], and the deterministic fabrication of QD-resonator structures[4,24,25]. To develop QD-based

---
[a)]pavel.avdienko@tum.de
[b)]jj.finley@tum.de

SPSs, the first requirement is to achieve control over the fundamental self-assembly process of InAs QDs with low density, allowing for tuning of the QD emission wavelength. Nevertheless, a wafer-scale strategy for epitaxial growth of QDs with low density and certain optical and structural properties is not yet completely realized. This arises from the steep onset of QD nucleation at a critical InAs layer thickness and a complex interrelation of the growth parameters (growth temperature, indium and arsenic flux, deposition time, growth interruption time, and overgrowth rate) that complicates the stable and reproducible formation of QDs with well-defined and uniform properties.

A gradient QD growth approach based on an intrinsic molecular beam non-uniformity along the sample surface enables a reliable way of controlling the surface density of QDs[4,11,13,19,26–30]. A comprehensive experimental study of the flux non-uniformity in MBE over stationary and rotating substrates has been conducted by *Wasilewski et al.*[31]. These authors showed that the layer thickness as a function of the position on the substrate without rotation can change in the range of $\pm 20\%$ along the diameter of a 2-inch wafer[29,31]. The critical InAs coverage of GaAs for the transition from the 2D to the 3D in Stranski-Krastanov (SK) growth mode ranges from 1.6 ML (measured by AFM) to 1.8 ML (measured by RHEED), as has been carefully studied by many groups[12,26,30,32–34]. Despite the universality of gradient growth, achieving scalable MBE growth of low-density InAs/Ga(In,Al)As QDs that emit within the targeted spectral range and that exhibit the required optical and structural properties demands precise control over the fundamental aspects of the self-organized processes and growth parameters across the entire substrate[33,35,36].

Here, we report the development of a molecular beam epitaxy (MBE) growth strategy for low-density self-assembled InAs/InGaAs QDs emitting in the O-band, suitable for the fabrication of SPSs. Our growth strategy relies on a gradient deposition of sufficiently large and In-rich InAs QDs grown in the sub-monolayer (ML) regime with a density modulations technique using a GaAs pattern defining layer (PDL)[26] and overgrown by an InGaAs strain reducing layer (SRL)[7,10,17]. Our method represents a universal approach for the MBE growth of low-density InAs/Ga(Al,In)As QDs with modulation of the areal density across the entire substrate. We correlate the gradient deposition of InAs QD ensembles by synchronizing the substrate rotation with the sub-ML deposition and investigate the impact of a growth temperature gradient across the substrate on the optical properties of QDs. Using the spatial distribution and the spectral emission properties of a large set ($>500$) of individual InAs/InGaAs dots characterized via hyperspectral imaging (HSI), we confirm that the developed growth strategy is universal and can be utilized for realizing quantum light sources for use in fiber-based quantum networks and integrated quantum photonic systems. Finally, we demonstrate electrical tunability of QD emission in the O-band and correlate results with atomic-scale QD structure and composition elucidated via aberration-corrected scanning transmission electron microscopy (STEM) and energy-dispersive X-ray (EDX) spectroscopy.

## II. EXPERIMENT

The InAs/Ga(In)As QD heterostructures were grown on undoped 2-inch GaAs(001) substrates using a Veeco Gen II MBE equipped with conventional effusion cells for group-III elements and a valved cracker cell for arsenic. All samples contain a buffer layer, 30 periods of a short-period AlAs (2 nm)/GaAs (2 nm) superlattice, followed by a 50–150 nm GaAs layer grown at a substrate temperature ($T_S$) of 600°C measured by a calibrated infrared IRCON pyrometer. Primarily, we used a $As_4$ beam equivalent pressure $P_{As}$(BEP) of $1 \times 10^{-5}$ Torr measured at the substrate position using a Bayard-Alpert ion gauge. Before the InAs deposition, we grew a 15 nm-thick GaAs pattern-defining layer (PDL) without rotation at $T_S = 580°C$, aligning the Ga-cell towards the secondary flat of the 2-inch GaAs(001) wafer[26]. Directly after the deposition of the PDL, we reduced $T_S$ for the QD growth, followed by a stabilization break. In addition to pyrometer measurements, the QD ($T_S$) was controlled using reflection high-energy electron diffraction (RHEED) by monitoring the transition of the GaAs surface reconstruction from $(2 \times 4)As$ to $c(4 \times 4)As$ under As flux. To grow QDs in SK growth mode, we deposited a gradient of 1.6–2.8 MLs of InAs at the substrate's center in the sub-ML regime with a sequence consisting of an InAs growth time $t_g = 3$ s and an As exposure pause time $t_p = 3$–51 s. The temperature $T_S$ was in the range of 480–550°C and the arsenic pressure was $P_{As}$(BEP) = $0.7 \times 10^{-5}$ Torr. After the QD formation, we reduced $T_S$ by 25–30°C during a 1–2-minute interruption under As supply. Subsequently, the InAs/GaAs QDs or InAs/InGaAs QDs were overgrown by 10-nm-thick GaAs or 7-nm-thick $In_{0.29}Ga_{0.71}As$ SRL and 2-nm-thick GaAs, respectively. After capping the InAs QDs, $T_S$ was linearly increased to 600°C and $P_{As}$(BEP) was set back to $1 \times 10^{-5}$ Torr for additional growth of 120–150 nm GaAs or $Al_{0.33}Ga_{0.67}As$ layers. Growth rates (r) of GaAs, AlAs, and InAs were $2.00 \pm 0.02$ Å/s, $1.00 \pm 0.01$ Å/s, and $0.150 \pm 0.003$ Å/s, respectively. The indium cell temperature remained constant during the InAs QDs and $In_{0.29}Ga_{0.71}As$ SRL growth.

Fig. 1 shows a schematic representation of the effusion cell position in the horizontal MBE growth chamber and the initial orientation of the wafer before the deposition of the GaAs PDL and the InAs/GaAs QDs. The standard dual filament effusion cell (C4 in Fig. 1) with a conical pyrolytic boron nitride (PBN) crucible (cone angle $\beta \approx 4°$) was used for gallium. The single filament indium effusion cell (C6 in Fig. 1) had a conical PBN crucible with $\beta \approx 8°$.

Photoluminescence mapping measurements were carried out using an experimental setup based on a He-flow cryostat under continuous wave (CW) non-resonant laser excitation with a wavelength of 660 nm. The excitation power density was varied in the range $\rho_{exc}$ up to 0.3 W/cm² with a spot diameter on the sample of $\sim 0.3$ mm. The luminescence was recorded using a spectrometer equipped with an InGaAs linear array detector. In these measurements, the PL mapping area of the 2-inch wafers in the He-flow cryostat is limited to 45 mm in diameter.

Structural measurements of the QDs were performed using atomic force microscopy (AFM) and STEM. Here, a Bruker Dimension Icon system was used to perform AFM measurements in tapping mode and a Titan Themis 60-300 STEM equipped with an aberration corrector for the probe-forming optics and a chemiSTEM EDX system was used for structural characterisation, operated at 300kV.

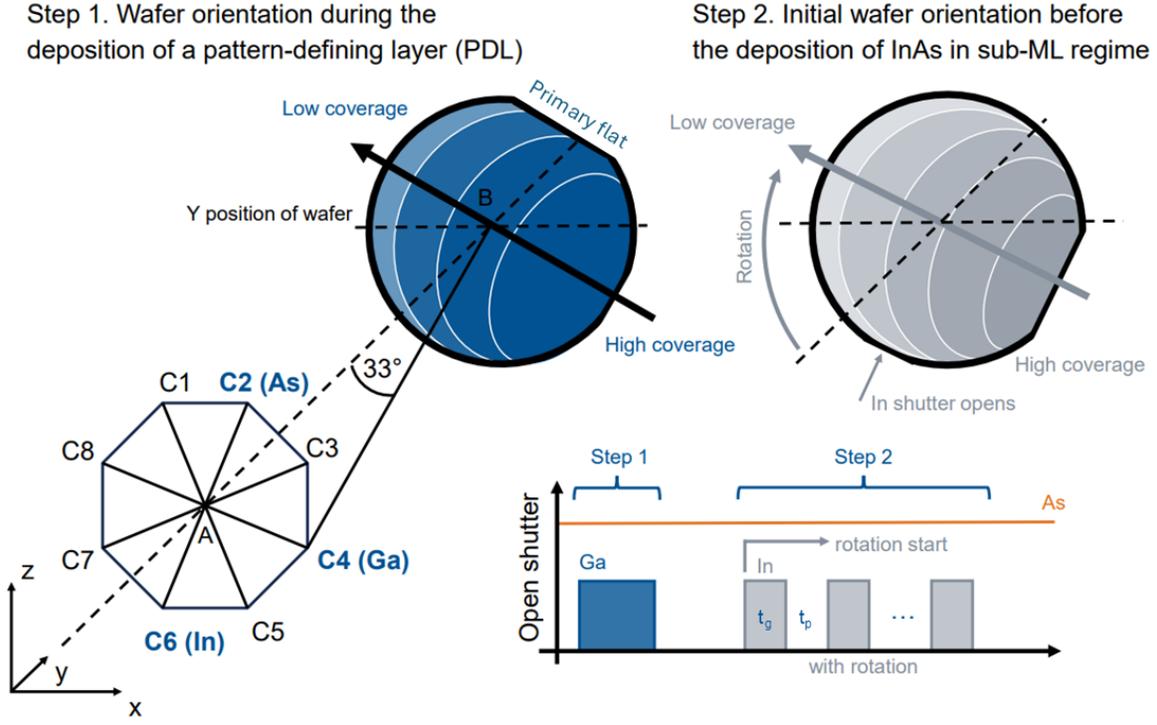

FIG. 1. Schematic of the initial cell-substrate geometry (xy is a horizontal plane) during the deposition of the GaAs PDL at $T_S = 580°C$ (Step 1) and initial orientation of the substrate before deposition of InAs in the sub-ML regime. With $t_g$ and As exposure break time $t_b$ (Step 2) at $T_S^{InAs}$, the shutter sequence for Ga, In and As cells employed during the PDL growth and the sub-ML InAs deposition cycles are synchronized when $t_g = T/2$ and $t_p = (2N + 1) \cdot t_g$, where T is a substrate rotation period and N are the natural numbers.

## III.  RESULTS AND DISCUSSION

### A.  Gradient growth of InAs QDs on the GaAs substrate

The key features of our developed growth strategy are the implementation of a PDL to modulate the surface roughness on the substrate and optimization of the sub-ML InAs deposition cycle sequence. Specifically, before InAs QD growth, we deposit a 15 nm-thick GaAs PDL without rotation, aligning the Ga-cell towards the secondary flat of the 2-inch GaAs(001) substrate (Step 1 Fig. 1). Immediately after the GaAs PDL deposition, the $T_S^{InAs}$ is ramped down from 580°C to 520°C before subsequent InAs deposition without excessive In desorption and preserving the GaAs PDL surface morphology. Optimization of the sub-ML InAs deposition cycle sequence was carried out considering that QD nucleation in SK growth mode can be described by the "superstress" parameter $\vartheta$, taking into account the case of unintentional In-Ga intermixing on the interface at $T_S^{InAs} > 480°C$[34]. This param-eter is defined as $\vartheta = (\theta/\theta_{eq} - 1)$, where $\theta$ is the coverage, and $\theta_{eq}$ is the quasi-equilibrium coverage[37]. Nucleation begins once the wetting layer (WL) reaches a critical coverage $\theta_{cr}$, at which point $\vartheta > 0$. The nucleation stages of the SK growth mode are divided into the following: 1. Formation of a wetting layer with $\theta_{eq}$; 2. As the coverage increases beyond $\theta_{eq}$, the layer becomes metastable but insufficient for QD nucleation; 3. When the deposited layer reaches $\theta_{cr}$, the nucleation barrier is minimized, while the nucleation rate is maximized.

The first step of the MBE growth optimization is the selection of the growth temperature $T_S$. Higher $T_S$ provides an increase in island volume due to the increased adatom diffusion coefficient and weaker influence of diffusion barriers[12]. Based on our RHEED pattern transition from 2D to 3D during InAs/GaAs QD growth at the substrate center with rotation and continuous deposition, the InAs desorption rates are $5 \times 10^{-4}$ and $2.5 \times 10^{-3}$ ML/s at $T_S = 530°C$ and 540°C, respectively. To minimize In desorption and maximize the diffusion length of In adatoms, all samples discussed in this paper were grown at $T_S = 520$–525°C, with an As $P_{As}$(BEP) $= 7 \times P_{As}^{cr}$(BEP) $= 7 \times 10^{-6}$ Torr, where $P_{As}^{cr}$(BEP) is the critical value for observing the (2×4)As reconstruction on the GaAs(001) surface by RHEED during GaAs growth at $T_S = 600°C$ with growth rate of $2.00 \pm 0.02$ Å/s.

Fig. 2 compares the spatial maps of the integrated PL in-tensity over the range 1000 - 1400 nm from the InAs/GaAs QDs as a false colour representation. These datasets were recorded at room-temperature from the following three samples: (i) a sample grown using standard gradient deposition without rotation and PDL (**S1**, fig. 2a), (ii) samples grown with GaAs PDL and sub-ML InAs deposition without rotation (**S2**, fig. 2b) and (iii) a sample with 45°-step deposition cycles (**S3**, fig. 2c). All samples were grown at the same $T_S^{InAs}$ = 520°C and $P_{As}$(BEP) = $1 \times 10^{-5}$ Torr. The InAs coverage consistently varied across the substrate, with higher coverage observed near the substrate's primary flat and lower coverage towards the opposite primary flat side by depositing from an inclined In effusion cell. Each map is individually normalized to the peak integrated PL intensity. These maps reveal the evolution of the PL intensity pattern, transitioning from the standard InAs gradient deposition without a GaAs PDL (Fig. 2a) to a stripe pattern with a clear modulation of InAs/GaAs QD emission along the horizontal direction when the GaAs PDL is grown (Fig. 2b and c). The integrated PL intensity distribution along the coverage gradient exhibits a non-monotonic behaviour, reflecting the evolution of the emission properties as a function of the InAs coverage and the corresponding density of InAs/GaAs QDs. Moving from the primary flat towards the top of the wafer, the nominal coverage of InAs presented in Fig. 2a ranges linearly from 2.2 MLs (bottom) to 1.57 MLs (top). The PL map in Fig. 2a indicates that as the InAs coverage increases, the PL intensity reaches a maximum value within the nominal coverage range of 1.8 to 2.1 MLs, and decreases in regions with coverage below 1.8 MLs and above 2.1 MLs. This behaviour is characteristic of the gradient growth mode, the continuous decrease in integrated PL intensity below 1.8 ML reflects a direct reduction in QD density, while the drop in PL intensity above 2.1 ML arises from unfavourable QD ripening into larger islands, which accumulate sufficient elastic energy for plastic relaxation of the InAs islands with non-radiative recombination centers.[7,10,12,38]

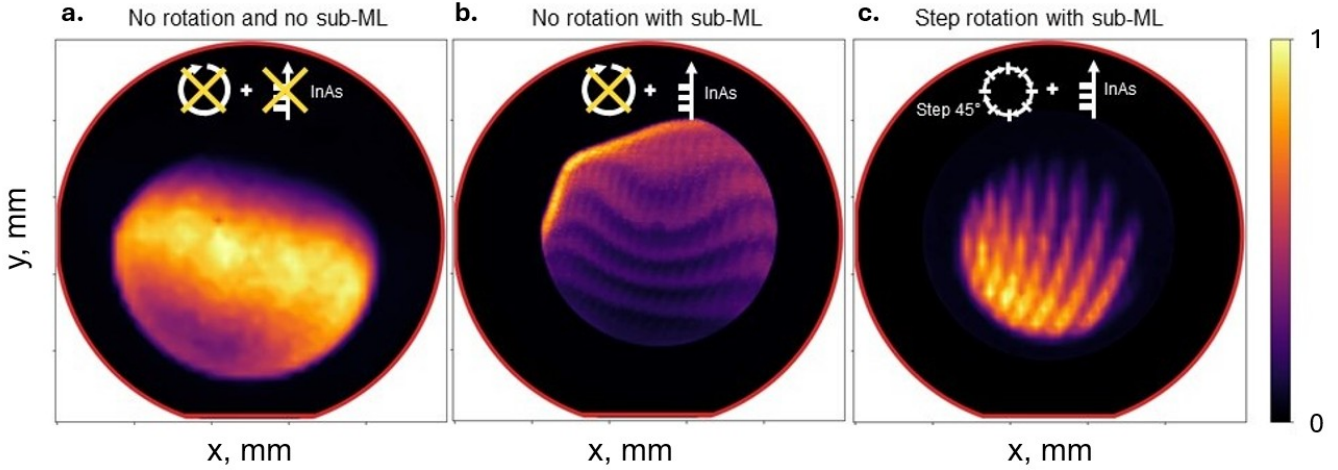

FIG. 2. False color maps of the integrated PL intensity from the InAs/GaAs QDs acquired at RT from the 2-inch GaAs(001) wafers. **a.** Sample **S1** is grown using standard gradient deposition for 40 seconds without rotation, or a PDL. Samples **S2 (b)** and S3 **(c)** are grown with a nominal 15-nm-thick GaAs PDL and a sub-ML InAs deposition. **S2** is grown without rotation by depositing 19 InAs cycles, each consisting of 3 s of growth followed by a 51 s break. **S3** is grown by depositing 12 InAs cycles, each consisting of 3 s of growth followed by a 5 s break, with the first 4 cycles without rotation, aligning the In-cell to the primary flat, while the remaining 8 cycles used 45° step rotation.

The growth of the Ga(In,Al)As PDL before the formation of the QDs can be used as a method to control the local dot density on the wafer[11,26,27]. The PDL grown without rotation in the layer-by-layer regime results in a surface roughness modulation across the entire substrate, consisting of fully and partially completed monolayers. This morphology is characterized by the presence of monolayer islands and atomic steps[26,31]. The atomic step density in the so-called rough region varies between approximately 11–12 µm$^{-2}$, whereas the regions classified as flat exhibit a much lower step density of about 3–5 µm$^{-2}$ (see Ref.[26]). When stress-driven nucleation is taken into account, the presence of a Ga(In,Al)As PDL increases local surface roughness, causing a preferential nucleation of InAs islands in the rough regions compared to the flat terraces due to the modification of the formation energy and decreases the nucleation barrier (free energy of island formation $\Delta G(rough) < \Delta G(flat)$)[37]. This modification means that nucleation in the presence of a surface step can proceed within the coverage interval $\theta_{eq} \leq \theta < \theta_{cr}$. Consequently, a roughness modulation on the substrate can lead to spatial ordering of QDs on the substrate, enhancing their nucleation by at least an order of magnitude[26]. Fig. 2b and 2c show PL maps of samples **S2** and **S3** grown with a nominal 15-nm-thick GaAs PDL. The nominal coverage gradient of sample **S2** ranges linearly from 3.3 MLs (bottom) to 2.4 MLs (top) in the vertical direction, as shown in Fig. 2b. QD growth on stationary substrates with the coverage greater than 2.4 MLs exhibits the same continuous decrease in PL integrated intensity as observed in conventional gradient deposition in Fig. 2a (close to primary flat). While it is well established that the formation of InAs/GaAs QDs is driven by elastic stress resulting from lattice mismatch, the kinetic growth mechanisms can vary signif-icantly depending on deposition conditions[12]. Implementing InAs sub-ML deposition with a long pause time on stationary substrates drastically modifies the PL intensity modulation of curvature fronts in the vertical direction. The $t_p$ of the sample **S2** was 51 s after every $t_g$ = 3 s InAs deposition cycles at T$_S$ = 520°C. The observed modulation in PL intensity can be attributed to the presence of a generation boundary, arising from the successive nucleation of multiple QD generations once the critical layer thickness $\theta_{cr}$ is locally reached.

We explain the generation boundary considering that QD nucleation occurs at a critical InAs coverage of $\theta_{cr} \approx 1.6$ ML[32,33]. This can be during InAs deposition once the coverage reaches $\theta_{cr}$ or during the interruptions between cycles as a result of kinetic processes. In regions of low QD density during the first phase of QD growth, and in the absence of direct In vapor impingement, QDs can grow in size via a metastable accumulation of adatoms on the WL surface, as well as through solid-state diffusion. In the latter case, the QDs grow at the expense of the WL itself[37] such that the residual WL becomes consistently thinner and less strained. Subsequent InAs sub-ML deposition primarily increase the size of the preexisting QDs, thereby reducing the probability of new QD nucleation across the low-density front. Simultaneously, the continued sub-ML deposition in regions without QDs drives the local coverage towards $\theta_{cr}$, initiating the nucleation of a second QD generation. This process repeats with subsequent cycles. Adjusting the cycle duration alters the spacing between QD generation fronts on the wafer surface. The curvature of the transition front observed in Fig. 2b arises from a heat-sinking effect in the MBE system that induces a temperature gradient across the wafer[35]. This temperature profile shifts the transition fronts in the direction of higher coverage (see Fig. A1 and A2 in the appendix). The longer the overall cycle time $t_{cycle} = t_g + t_p$, and the stronger the radial temperature gradient from the wafer center to its edge, the more pronounced this curvature becomes, primarily due to enhanced In desorption.

Fig. 2c presents the PL map of sample **S3**, which was grown by depositing 12 InAs cycles. Each cycle consisted of $t_g$ = 3 s of growth followed by $t_p$ = 5 s. The first 4 cycles were grown without rotation, aligning the In-cell to the primary flat. The remaining 8 cycles employed a 45° step rotation to maximize $\theta$ in the central area, while also slightly shifting the maximum overlap towards the primary flat. The nominal $\theta$ gradient of sample **S3** ranges linearly from 1.8 ML at the center to 1.5 ML at the edge. The clear PL intensity modulation in Fig. 2c is a direct consequence of the roughness modulation in GaAs PDL[26].

Substrate rotation during growth improves not only the uniformity of the thickness profile[11,31] but also provides an effective technique to control the InAs coverage $\theta$ and QD density in the sub-ML regime, similar to step deposition applied for sample **S3**. Synchronizing the InAs sub-ML deposition times $t_g$ and $t_p$ with the substrate rotation period $T$ enhances both the uniformity and reproducibility of QD growth. Two particular synchronized sub-ML InAs deposition sequences can be defined when $t_g = T/2$ and $t_p$ as $2N \cdot t_g$ for nominally uniform coverage or $(2N + 1) \cdot t_g$ for gradient, where N is a nat-ural number. Our method to achieve a $\theta$ gradient in this work involves synchronizing $t_p = (2N + 1) \cdot t_g$ with N = 2 for all further discussed samples. This additionally results in a more convenient nucleation control via RHEED. The influence of the temperature gradient on the PL map pattern for such a gradient is presented in Fig. A2 of the Appendix. We note that a coverage gradient can also be achieved using a more complex sequence of deposition cycles that accounts for the non-uniformity of the In flux due to the specific cell geometry, the relative configuration of the In and As cells, and the temperature gradient on the substrate. For instance, a combination of cycles with and without rotation, along with close to uniform coverage synchronized with rotation, can be applied to control the QD nucleation. The PL map pattern evolution observed in Fig. 2 confirms that the PDL technique can be universally applied in conventional MBE growth on (001) surfaces regardless of the used MBE system. Moreover, it likely holds promise to be extended to the epitaxial growth of other materials systems involving strain-driven self-assembly of III-V, II-VI, and group-IV QDs.

### B. Structural and optical properties of InAs/Ga(In)As QDs

In the past, a few MBE growth strategies have been successfully developed for the growth of heterostructures with InAs QDs overgrown by an In(Ga,Al)As SRL with QD densities below 1 QD per µm². These enabled the development of GaAs-based SPSs emitting in the O-band around 1300 nm: using ultralow InAs growth rate[10] and formation of a bimodal distribution with many taller dots[7]. The overgrowth of InAs QDs grown on GaAs surface by an In(Ga,Al)As SRL with nominal thickness 5–8 nm and In-content 15–30% results in deeper confinement of charge carriers in the InAs/In(Ga,Al)As QDs due to the complex interplay between increased QD size, changes in the QD aspect ratio, and strain reduction provided by the In(Ga,Al)As capping layer. Typically, the ground state QD emission shift to longer wavelengths does not exceed 150 meV[7,10,15,17,39,40], which restricts optimizing the MBE growth conditions: initial InAs QDs with low density overgrown by GaAs should be In-rich and large enough to emit around 1200 nm (1033 meV) at LT. This ensures that the subsequent overgrowth by In(Ga,Al)As will lead to the redshift of the QD emission into the O-band around 1300 nm (954 meV) (see Figs. 3a and 3b).

Considering these conditions, our developed MBE growth strategy of the InAs/InGaAs QDs has been divided into two parts: (i) InAs coverage control using a shutter-synchronized sub-ML deposition in gradient mode to obtain low-density InAs QDs on the GaAs surface; and (ii) optimization of the overgrowth step with 7-nm-thick In$_{0.29}$Ga$_{0.71}$As SRL. While conventional gradient QD growth is a well-established method for controlling QD density[30], a reduction in InAs coverage is typically accompanied by a blue shift of the ground-state emission at LT (see Fig. A4 in the Appendix), consistent with earlier studies attributing this behavior to reduced QD size[36,41]. As we show in Fig. A4 of the appendix, the blue shift of the QD emission in the low coverage region can be overcome using coverage control in shutter synchronized InAs sub-ML deposition mode, which enables precise control of the InAs coverage. All samples discussed in the following were grown using 15 InAs sub-ML cycles synchronized with substrate rotation, with $t_g = 3$ s and $t_p = 15$ s times at a rotation speed of 10 RPM.

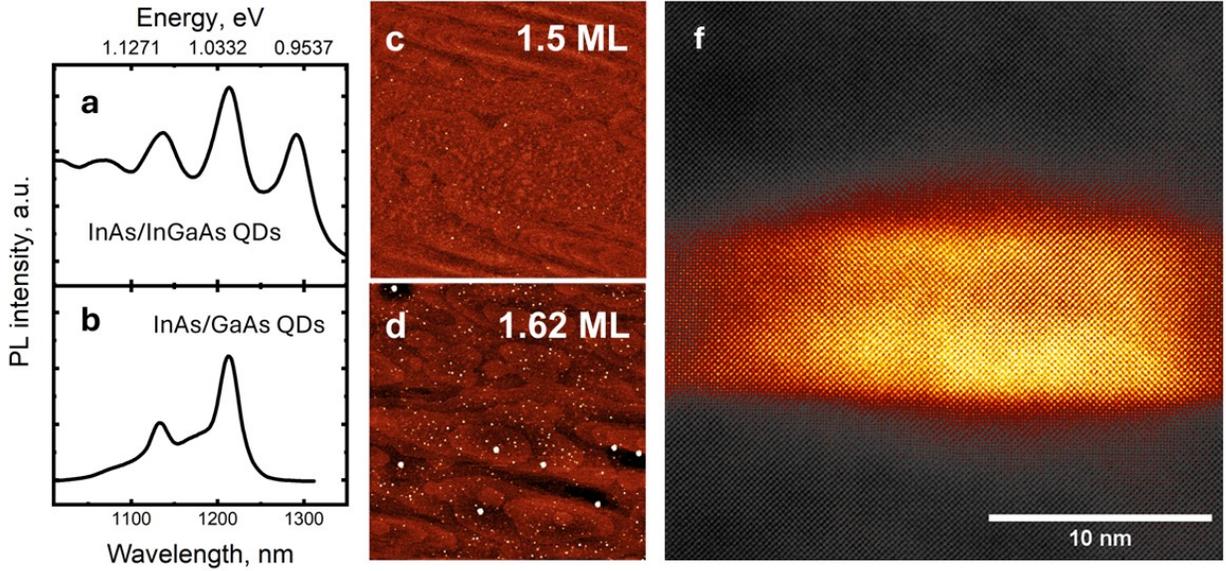

FIG. 3. **a** and **b**. Typical ensemble state-filling photoluminescence spectra of InAs/InGaAs and InAs/GaAs QDs measured in the low-density regions at 10 K with an excitation power density of 0.3 W/cm$^2$, respectively. **c** and **d**. 2×2 μm$^2$ AFM images for InAs coverage 1.5 ML and 1.62 ML along the gradient direction, respectively. **f**. False color cross-sectional high-angle annular dark-field scanning transmission electron microscopy (HAADF-STEM) image of a single InAs/InGaAs quantum dot grown at 520°C and overgrown at 490°C.

The subsequent overgrowth of InAs QDs by GaAs or InGaAs SRL inevitably modifies both the QD morphology and In content, accompanied by In–Ga intermixing. Even prior to the capping, nominally pure InAs QDs exhibit a vertical compositional gradient, with the Ga content near the base reaching values exceeding 60%[42]. To suppress additional In–Ga intermixing during overgrowth, we therefore reduced the overgrowth temperature of InAs QDs overgrown with GaAs. Fig. 4 shows the evolution of PL spectra for InAs/GaAs QDs overgrown at $T_S$ between 490°C (top panel) and 520°C (bottom panel). The spectra were acquired in the regions with QD densities as low as 10 QDs/μm$^2$. A clear trend is observed: increasing the overgrowth temperature reduces the localization energy, typically attributed to either a reduction in QD volume or enhanced In-Ga intermixing[12]. During the capping process, the island height is reduced as the GaAs overlayer is deposited, driving a thermodynamically favorable flattening of the QDs. This process broadens the lateral flanks of the islands[42] and substantially alters the QD aspect ratio, thereby modifying both the effective QD volume and the indium composition. As a consequence, the optical and struc-tural properties of nominally pure InAs QDs are strongly affected. A similar evolution of morphology and composition is expected during overgrowth with an InGaAs SRL. Based on these observations, InAs/GaAs QDs emitting around 1200 nm at LT were first obtained at a reduced substrate temperature of $T_S = 490$°C. Subsequent capping with the InGaAs SRL was therefore carried out at the same temperature to minimize In–Ga intermixing. The In flux was kept constant throughout both InAs QD growth and SRL deposition, followed by a 2-nm-thick GaAs cap. After completion of the SRL overgrowth, the $T_S$ was ramped back to 600°C for continued GaAs growth. The ripening interruption time after the final InAs deposition cycle was fixed at 90 s.

A pronounced (around 80 meV) ground-state emission redshift of InAs/InGaAs compared to InAs/GaAs QDs is presented in Figs. 3a and 3b, measured at 10K with an excitation power density of 0.3 W/cm$^2$. Figure 3f presents a high-angle annular dark-field scanning transmission electron microscopy (HAADF-STEM) image of a representative InAs/InGaAs QD exhibiting a characteristic lens-like morphology. From the HAADF–STEM analysis, the QD aspect ratio $H/L$, where $H$ denotes the QD height and $L$ the base diameter, is found to be approximately 0.2. This value is smaller than the aspect ratios of 0.3–0.37 typically obtained from AFM measurements in Fig. 3d, reflecting the strong modification of the QD shape during capping. As demonstrated in Ref.[40], the excitonic emission redshift in the InAs/InGaAs QD system can reach up to ~160 meV and depends strongly on the final QD shape and In composition, the $H/L$ ratio, and the strain redistribution introduced by the SRL. In particular, the largest redshifts are predicted for aspect ratios close to 0.2, resulting from the combined effects of quantum confinement and the interplay between hydrostatic and deviatoric components of the elastic strain tensor[40]. Based on these calculations, we might estimate that strain-field redistribution contributes approximately 50–60 meV to the redshift for an InGaAs SRL, while the combined effects of increased QD volume and a reduced $H/L$ aspect ratio account for an additional 30–50 meV. Together, these mechanisms explain the typical QD emission redshift of ~80–100 meV observed in our experiments (see Figs. 3a,b). To further assess the strain and composition, an aligned series of HAADF-STEM images was used to minimize scan distortions, revealing a strain along growth direction of approximately 3–4% in the InGaAs SRL. By elasticity theory applied to InGaAs/GaAs systems[43], this translates to In contents of $x = 0.29 \pm 0.03$ within the SRL, which we used to calibrate the HAADF-STEM intensity of the same SRL region. According to Grillo et al.[44], the dependence of the STEM signal on x is approximately linear, from which the average indium content in the interaction volume containing the QD could be derived. Since the QD extension along <100> direction can be assumed isotropic, the extent of the QD along beam direction is assumed equal to its lateral extension. By measuring the specimen thickness to be 82 nm via comparison of experimental with simulated position-averaged convergent beam electron diffraction patterns, and taking the broadening of the convergent electron beam in the specimen into account, the increase of the HAADF signal in the QD with respect to the SRL could be assigned to a QD with an In content of $x_{min} = 0.41$ and $x_{max} = 0.49$.

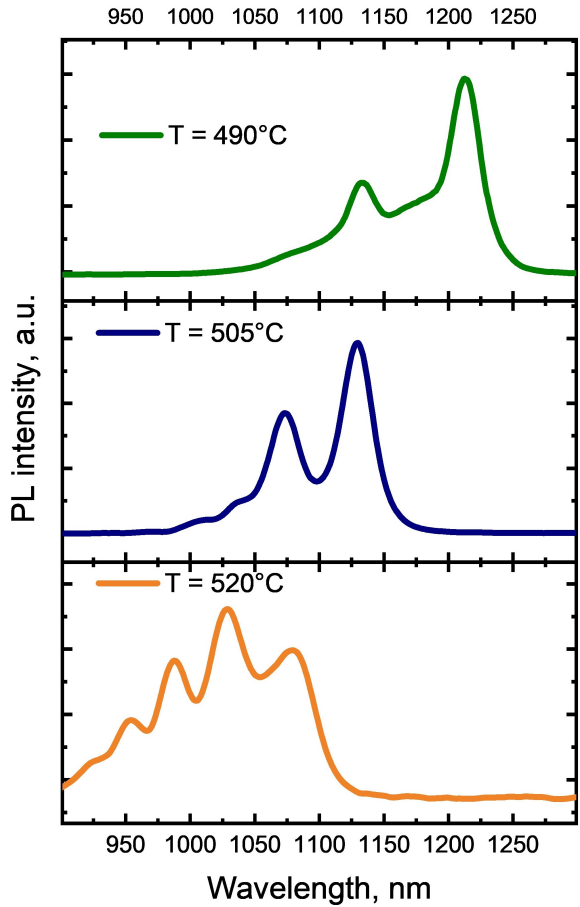

FIG. 4. PL spectra for InAs/GaAs QDs overgrown at $T_S = 490°C$ (top panel), 505°C (middle panel) to 520°C (bottom panel), respectively. All PL spectra are acquired at T = 10 K in the regions where QD density is as low as 10 QDs/μm².

We now demonstrate how these findings can be combined to realize universally scalable MBE growth of low-density O-band quantum dots. Fig. 5a shows a false-color map of the integrated PL intensity of a fourth sample, denoted **S4**, containing InAs/InGaAs QDs grown at $T_S = 520°C$ and overgrown at $T_S = 490°C$. Figure 5b shows the corresponding map of the ground-state emission peak of the same sample extracted from PL spectra measured at 10 K. Focusing on the investigation of the structural and optical properties of the InAs/InGaAs QDs grown on GaAs surface with $\theta < 1.8$ ML (region III in Fig. 5). For AFM measurements of surface QDs, rapid cooling was applied immediately after the completion of the InAs deposition cycles. Figs. 3c and 3d compare AFM images recorded along the gradient thickness in regions corresponding to nominal $\theta \approx 1.5$ and $\approx 1.62$ ML of InAs, respectively. The AFM images in region III reveal the presence of two distinct InAs/GaAs QD populations: small and large QDs. The small QDs have heights ranging from 1 to 5 nm with a base length of around 15-20 nm, while the large QDs exhibit heights between 12 and 17 nm and base lengths around 40–45 nm. The density of the large QDs, which are responsible for the origin of the PL signal, decreases as the InAs coverage decreases (see Fig. A4 in the Appendix). Small QDs are primarily observed in the coverage range of 1.45–1.65 MLs, while large QDs begin to appear and increase in density, reaching approximately 10 QDs/μm² as the coverage approaches 1.65–1.7 ML. Beyond 1.6 ML, the density of small QDs diminishes, while the population of large QDs continues to increase (see Fig. A5). Our AFM data confirm that the density evolution of both small and large QDs as a function of InAs coverage is consistent with the observations of Placidi et al.[33] and further verify the dominant role of QD nucleation in rough regions of the GaAs PDL[26]. The evolution of two QD populations may be quantitatively explained considering that in the initial step with $\theta < 1.6$ ML, the 3D islands are separated by an average distance $d$, much larger than the In diffusion length $l$. Under these conditions, no cation exchange occurs between different islands, and island growth proceeds independently through the expense of WL and available In adatoms affecting both island types. With increasing $\theta > 1.6$ ML, the island density rises drastically[32] such that $d \propto 2l$. In this regime, cations are sufficiently close to migrate towards the thermodynamically favorable size under the applied MBE growth conditions. This preferential migration diminishes the populations of small QD and gradually narrows the island size distribution. This confirms that, at sufficiently low InAs coverage, the QD size evo-lution is primarily governed by kinetic mechanisms. Regardless of the QD generation, nucleation proceeds more rapidly in rough regions of the GaAs PDL than on flat regions, resulting in higher QD densities along rough PDL trajectories and, consequently, stronger integrated PL intensity. One trajectory along the rough region is highlighted by orange (Fig. 5a) and red (Fig. 5b) lines for clarity.

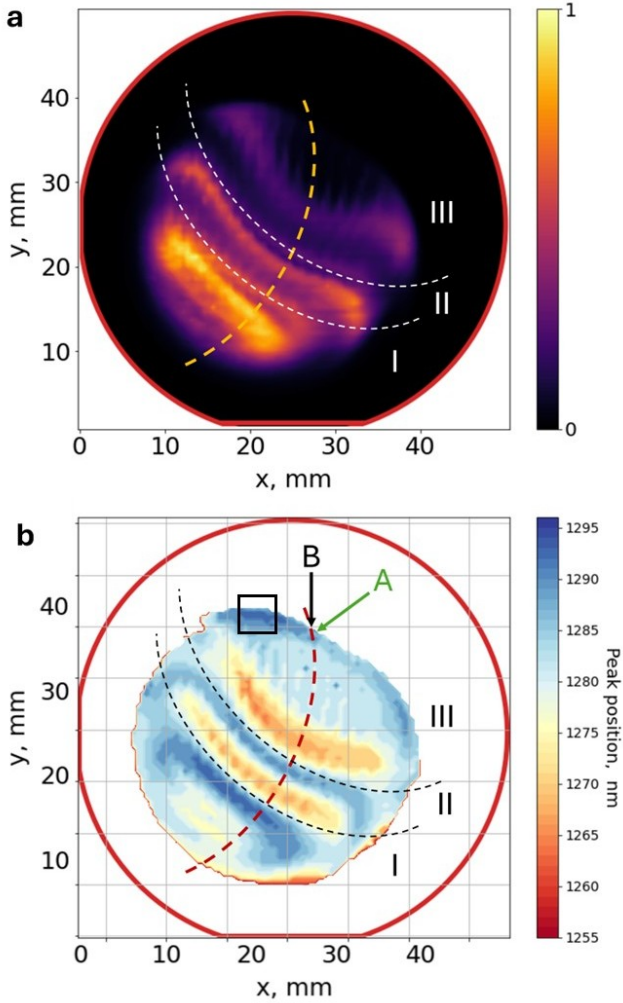

FIG. 5. **a.** False color integrated PL intensity map of the sample **S4** with the InAs/InGaAs QDs. **b.** InAs/InGaAs QD ground state peak wavelength map. The PL map was measured at 10 K with an excitation power density of 0.3 W/cm². The single trajectory along the rough region is highlighted by orange (**a**) and red (**b**) lines for clarity.

Fig. 5b shows a slight blueshift of the PL emission along the rough PDL trajectories of the last QD generation (region III) in the absence of In vapor impingement, which might be caused by cation exchange between the islands at $\theta > 1.6$ ML and $d \propto 2l$. However, when the coverage $\theta > 1.7$ ML under the sub-ML deposition (I and II generation in Fig. 5), direct In vapor impingement becomes the dominant factor in modifying the QD size, and In content of QDs and WL before overgrowth. The preexisting QDs become larger, which results in a pronounced redshift of QD emission at the generation boundary. This process repeats during deposition in the regions of subsequent QD generations. The combination of the gradient growth approach and synchronized sub-ML deposition developed here provides an additional degree of freedom to control the surface density and size of the QDs. Moreover, our developed approach inherently incorporates the previously mentioned bimodal distribution growth modes[7] in regions I and II for achieving long-wavelength emission for individual large QDs, particularly in cases where more than two QD generations are present on the substrate. Large InAs QDs at nominal coverage ∼1.6 ML exhibit sizes comparable to those achieved using ultralow InAs growth rates[10] or via bimodal distribution growth modes[7]. In addition, we note that the developed growth approach can be adapted to tailor the emission wavelength of InAs/GaAs quantum dots in the shorter < 1200 nm spectral range by employing the so-called indium-flush step[28,30].

### C. Micro-PL spectroscopy of InAs/InGaAs QDs

We continue to validate the results of the AFM and PL mapping experiments in a low-density gradient region with a maximum nominal InAs coverage of up to 1.65 ML (corresponding to the last QD generation III) by performing micro-PL (μPL) measurements at 4 K. The measurements were carried out using a confocal microscopy setup with CW excitation at 900 nm (above the SRL) and an apochromatic objective with a numerical aperture of 0.81. The QD emission was detected using a spectrometer with a focal length of 750 mm, equipped with a 1200 lines mm⁻¹ grating and an InGaAs linear array detector. The measurements were performed in a region of sample **S4** located in the black area indicated in Fig. 5b. Fig. 6a shows the normalized excitation power dependence of the μPL spectra of a representative single InAs/InGaAs QD. The saturation power of the transition labeled CX, emitting at 957.924 meV, is determined to be $P_{sat} = 10$ μW under 900 nm CW excitation. For the transition lines observed in non-resonantly excited QDs, a typical full width at half maximum (FWHM) of 20–30 μeV (5-7.3 GHz) is extracted. The observed linewidth is limited by the spectral resolution of the μPL setup. The identification of the excitonic states was based on the excitation power dependence of the different charge configurations of the same emitter. The μPL spectrum of an individual InAs/InGaAs QD exhibits a characteristic emission pattern that might be assigned to the neutral exciton (X), biexciton (XX), and charged exciton (CX) transitions. The X and CX show linear dependences of the μPL intensity on the excitation power density, with fitted slopes of $0.95 \pm 0.05$ and $1 \pm 0.05$, respectively, while the biexciton (XX) exhibits a quadratic dependence with a slope of $1.9 \pm 0.1$. To verify the single-photon emission characteristics of the InAs/InGaAs QDs, second-order intensity autocorrelation measurements were performed under CW excitation using the CX transition line. Typical results are presented in Fig. 6b. A pronounced suppression of multiphoton events at zero time delay, $g^{(2)}(0)$, is observed, confirming the single-photon nature of the emission. From a fit to the autocorrelation data using $g^2(\tau) = 1 - A \exp(-|\tau|/\tau_a)$, we extract a value of $g^{(2)}(0) = 0.020 \pm 0.014$, $A = 0.980 \pm 0.014$ and $\tau_a = 0.75 \pm 0.01$ ns.

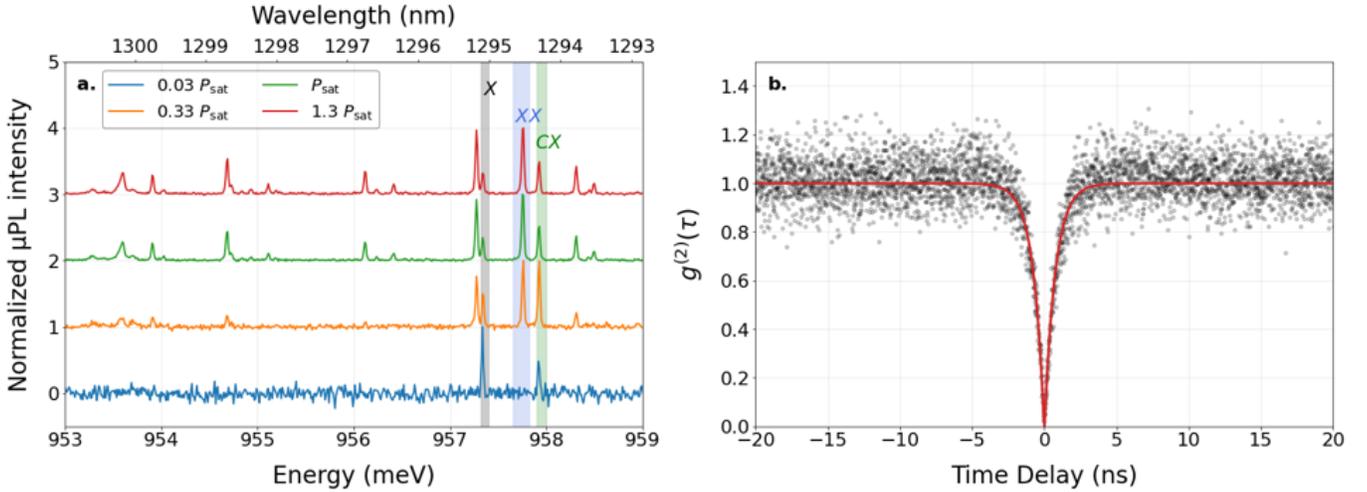

FIG. 6. **a.** Normalized excitation power dependence of μPL spectra of a representative single InAs/InGaAs quantum dot measured at 4 K. **b.** Second-order autocorrelation measurements of the CX transition lines under CW excitation at 900 nm with a saturation power $P_{sat} = 10$ μW.

To validate the presence of the PDL, we used a μPL scanning hyperspectral imaging technique (HSI) to characterize and statistically analyze the emission properties of individual InAs/InGaAs QDs as described in detail by Buchinger et al.[45]. Sample **S5** was grown under growth conditions similar to those used for sample **S4**. The measurement routine of sample **S5** was divided into two parts: a single snapshot image and a hyperspectral imaging scan (see Fig. 7a). An alignment grid was placed on the sample for navigation. The sample was excited with a 730 nm CW laser diode in a closed-cycle cryostat at a base temperature of T = 4 K. In addition, for imaging, the sample surface was illuminated with a 1100 nm LED. Fig. 7a presents a typical PL spectrum of an ensemble InAs/InGaAs QDs at T = 10 K from regions with QD densities as low as 1 QD/μm², together with the results of HSI measurements and subsequent statistical analysis. The emission wavelength distribution of 505 individual InAs/InGaAs QDs was obtained from six $20 \times 20$ μm² HSI areas: three in nominally flat area A (plotted in red) and rough B (plotted in blue) PDL regions similar to marked in Fig. 5b) with QD density as low as 0.15 QD/μm² (Fig. 7a). A peak emission histogram of more than 500 QDs measured in both A and B PDL regions illustrates a pronounced maximum around a peak wavelength near 1310 nm with an asymmetric short-wavelength tail. The asymmetry of the peak emission histogram can be attributed to two main reasons: The presence of QDs with a smaller size and/or reduced indium composition emitting at shorter wavelengths than the O-band range (< 1260 nm), predominantly localized in regions with greater roughness. This is in good agreement with the PL spectrum shown in Fig. 7a. For the so-called p-shell emission peak, an additional shoulder around 1200 nm can be seen, which corresponds to the presence of QDs ground state whose transition is in the region of $\leq 1300$ nm. We note that the emission wavelength of individual InAs/InGaAs QDs can vary significantly from that of neighboring dots, particularly in regions affected by surface roughness. Moreover, the majority of QDs emitting in the O-band were found in the area labelled A on Fig. 5b, as confirmed by PL mapping. The homogeneity of the QD ensemble emission can be significantly improved by employing the synchronization approach described earlier[26].

### D. Electrically tunable device

As can be seen in Fig. 7a, the fabrication of efficient SPSs calls for the site-selective definition of nanophotonic structures that are resonant with the QD emission[1]. This challenge can be solved using in-situ photo- or electron beam lithography or marker-based PL imaging, followed by the lithography steps[45]. Precisely controlling the detuning between the QD and the resonator wavelength can be achieved by adjusting the temperature at the expense of enhanced phonon-mediated decoherence. Hence, a better solution to fine-tune the detuning is tuning the QD transition through the quantum confined DC Stark effect (QCSE)[1,46]. Therefore, we characterize the QCSE tuning range of the grown InAs/InGaAs QDs. This information is also very relevant for the design of QD-molecule-based heterostructures that use vertically correlated pairs of dots[4]. To this end, we embedded a single layer of self-assembled QDs, similar to sample **S4**, within the intrinsic region of an n-i-Schottky photodiode. The InAs/InGaAs QDs were positioned at the center of a $\lambda$ cavity, 35 nm above a 50-nm-thick n-GaAs back contact doped to $5 \times 10^{18}$ cm⁻³, followed by a 164-nm-thick GaAs layer, a 40-nm short-period AlAs/GaAs (3 nm / 1 nm) superlattice serving as a current-blocking layer, and capping with 9 nm of GaAs. The structure also incorporated a bottom DBR mirror consisting of 14 GaAs/AlAs pairs to enhance photon collection efficiency. Standard semitransparent Ti/Au top and AuGeNi-based bottom contacts were used to gate the photodiode.

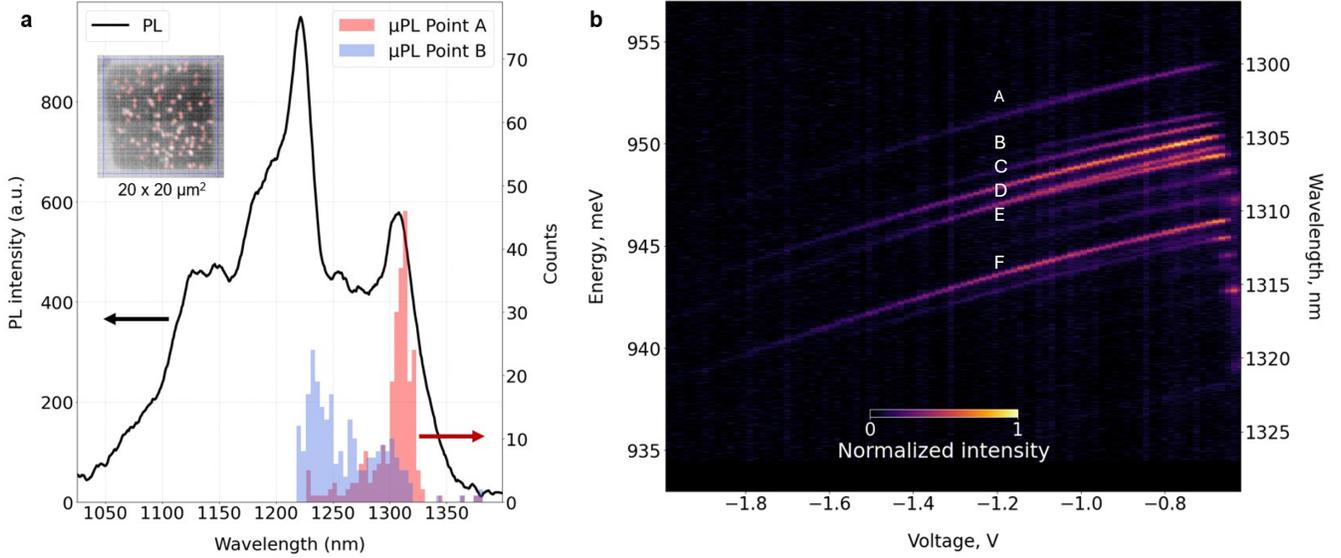

FIG. 7. **a.** Typical sample **S5** PL spectrum of ensemble InAs/InGaAs QDs acquired at T = 10 K in the regions where QD density is as low as 1 QD/μm². The emission wavelength distribution of 505 individual InAs/InGaAs QDs acquired using HSI areas in the region A and B marked in Fig. 5b. The inset shows an example of a 20×20 μm² studied area with QD density as low as $10^7$ cm⁻². **b.** Typical μPL spectra of single InAs/InGaAs QDs as a function of applied electric field acquired at T = 4 K in the regions where QD density is as low as 1-2 QDs/μm² under 900 nm CW non-resonant excitation.

Fig. 7b. demonstrates a typical PL spectrum of nominally single InAs/InGaAs QDs as a function of applied electric field acquired at T = 4 K in the region where the QD density is as low as 1–2 QDs/μm². The various QD emission peaks, labeled A-F, shift to lower energy with increasing field. For voltages below -1.8 V, no PL signal is observed due to field ionization of excitonic states. As the electric field reduces, the energy of excitonic states shifts due to the QCSE. The shift is well described by

$$E(F) = E_0 - p \cdot F - \beta \cdot F^2,$$

where $E_0$ denotes the transition energy in the absence of an external field, $p$ is the static exciton dipole moment, and $\beta$ is the polarizability[12,46,47]. The electric field **F** is a function of the applied voltage $V$ and the thickness of the region $D$. It can be expressed as

$$F = \frac{V - V_0}{D},$$

A parabolic fit of the Stark-shifted A-F lines presented in Fig. 6b yields $p/e$ in the range of 0.3–0.4 nm and $\beta \approx$ 0.6–0.7 μeV × kV⁻² × cm². The quantity $p$ is interpreted as a spatial separation between the centers of the electron and hole envelope wave functions in the absence of an external electric field arising from the spatial gradient of In composition within the QD. Typically, due to the strain profile of the QD, the indium concentration increases from the base toward the apex and from the lateral flanks toward the centre of the dot[42]. As a result, the hole wave function is preferentially situated in the indium-rich upper region, whereas the less localized electron is positioned closer to the dot center[46]. The observed positive sign of the $p$ at $F = 0$ confirms this interpretation for the InAs/InGaAs QDs grown and studied in this work. In contrast, a negative $p$ is typically expected for QDs with a homogeneous In composition[46]. This is consistent with the behavior of InAs/GaAs QDs, although the obtained $p$ in InAs/InGaAs QDs is approximately a factor of two smaller than typical values reported for InAs/GaAs QDs grown without a flush step[12,46], which may additionally confirm that the above mentioned SRL capping carried out at $T_S$ = 490°C leads to the partial suppression of In–Ga intermixing and strain-driven In migration away from the QDs. As a result, the effective carrier confinement is stronger than one would expect based on the increase in QD volume. The parameter $\beta$ is most sensitive to the dot height. In the vertical direction $\beta$ increases rapidly with QD height, approximately following a scaling as $\beta \propto h_z^4$ (considering quantum well-like scaling)[48]. Taking into account the typical QD heights of $h_{1300\,nm}^{QD}$ = 5–6 nm and $h_{930\,nm}^{QD}$ = 2.5–3 nm, the ratio of polarizabilities can be estimated as $\beta_{1300\,nm}/\beta_{930\,nm} \approx$ 7–16. Experimentally observed polarizability of In(Ga)As/GaAs QDs, as carefully studied by many authors[12,46], typically lies in the range $\beta$ = 0.1–0.4 μeV × kV⁻² × cm². These values are approximately 1.5 to 7 times smaller than the polarizabilities studied InAs/InGaAs QDs in our work. Consequently, InAs/InGaAs QDs exhibit a stronger parabolic Stark shift compared to In(Ga)As/GaAs QDs under an applied electric field. The combination of deeper exciton localization up to 550 meV relative to GaAs, a $p/e$ compared with InAs/GaAs flush QDs, and a large polarizability, together with the MBE growth approach developed in this work, makes InAs/InGaAs

QDs promising candidates for integration into the active region of electrically controlled single-photon sources (SPSs) and for the development of more complex heterostructures based on QD-molecules emitting in the telecom O-band[4].

## IV. CONCLUSION

In summary, we demonstrated a scalable MBE growth strategy to achieve high-quality low-density InAs/InGaAs QDs on standard GaAs(001) substrates with spatial QD density modulation emitting in the telecom O-band at cryogenic temperatures. By combining sub-ML gradient deposition with the InGaAs SRL approach, the QD emission wavelength was effectively extended while maintaining controlled density. Synchronization of substrate rotation with InAs sub-ML deposition, together with roughness modulation, proved highly effective in spatially controlling QD density, enabling regions with densities below $10^8$ cm$^{-2}$. PL mapping and HSI techniques is powerfull combination for optimization and control of high-quality growth of InAs/Ga(In)As QDs. Atomic-scale structure and composition are clarified via aberration-corrected STEM and EDX spectroscopy. PL mapping and hyperspectral imaging of QDs confirmed the structural and optical quality of the ensemble and validated the universality of the proposed method. We demonstrate the single-photon emission of the InAs/InGaAs QDs under CW excitation using the CX transition line with $g^{(2)}(0) = 0.020 \pm 0.014$. The results highlight that the developed approach can be readily implemented in conventional MBE systems on (001) surfaces and offers a promising pathway toward scalable fabrication of electrically tunable single-photon and entangled-photon devices for quantum photonic applications.

## ACKNOWLEDGMENTS

We thank Beatrice Costa from the Technical University of Munich for her assistance in setting up the optical setup. We thank F. Dushimineza from the Ludwig-Maximilians-University for assistance in HAADF-STEM measurements. We gratefully acknowledge the Deutsche Forschungsgemeinschaft (DFG) via projects DIP (FI 947/6-1), MU4215/4-1 (CNLG), INST 95/1220-1 (MQCL) and INST 95/1654-1 (PQET), Germany's Excellence Strategy (MCQST, EXC-2111, 390814868) as well as the Bavarian Ministry of Economic Affairs (StMWi) via projects 6GQT and Munich Quantum Valley via NeQuS. In addition, we gratefully acknowledge the German Federal Ministry of Research, Technology and Space (BMFTR) via the projects PhotonQ (FKZ 13N15760 and 13N15759), QR. N (FKZ 16KIS2197 and 16KIS2209), and 6G-life. A.L. acknowledges support by the BMFTR funded projects QTRAIN No. 13N17328, EQSOTIC No. 16KIS2061, and QR.N No. 16KIS2200, and the DFG funded project EXC ML4Q LU 2004/1. T.H.-L. acknowledges funding from the BMFTR via Qecs (FKZ 13N16272). A.T.P. acknowledges funding from the BMFRT via Ferro35 (FKZ 13N17641). K.M.-C. acknowledges funding from the DFG MU3660/4-1 and funding from the European Research Council within the Horizon Europe innovation funding programme under Grant Agreement 101118656 (ERC Synergy project 4D-BioSTEM).

## DATA AVAILABILITY STATEMENT

The data that support the findings of this study are available from the corresponding author upon reasonable request.

## APPENDIX

The substrate temperature value is a critical process parameter in MBE growth. The temperature gradient across a wafer can lead to significant variations in the structural and optical properties of heterostructures grown by MBE. However, accurately measuring growth temperatures is not easy and depends on the equipment used in certain MBE setups. Moreover, accurately measuring the temperature gradient is more complicated and is mostly determined by the heater and substrate holder design. PL mapping is a powerful tool for optimizing a growth process, especially for the gradient growth mode of QDs or QWs.

In our MBE system, the substrate temperature is determined using a calibrated optical pyrometer, which records the temperature from the central region of the substrates. Based on a PL map analysis of samples **S1** and **S2** and a simulation of the InAs coverage on the substrate surface grown without rotation, we find that the region of reliable temperature is limited to the inner diameter of around 4 cm of the 2-inch substrate. Therefore, all further optimization steps and data analysis were carried out with this limitation of the temperature gradient in mind. The molybdenum wafer holder is typically up to 25°C hotter than the GaAs wafer once the system stabilizes at the growth temperature. Consequently, an additional heating of the substrate edges occurs from the holder through contact at the ledge[49]. However, it should be emphasized that when the diameter of the heater is smaller than the substrate (see Fig. A3 in the appendix), the wafer edges can be 30–40°C cooler, reducing the reliable temperature of the inner substrate diameter (see Fig. 2). The exact temperature distribution is influenced by the complex thermal coupling among the substrate, backing rings, platen, and manipulator heater, which is specific to the configuration of the employed system. As shown in Ref.[35], the temperature profile of the substrate may impact the structural and optical properties of the low-density InAs/GaAs QDs, depending on the location of the low-density QD region during nucleation, ripening, and overgrowth.

Fig. A2 shows an evolution of the PL map pattern of InAs/InGaAs QDs grown at different $T_S$ and the same synchronization parameters $t_g = T/2 = 3$ s and $t_p = 5 \times T/2 = 15s$.

A PL spectra analysis of the samples grown using a conventional gradient deposition without substrate rotation is presented in Fig. A4.

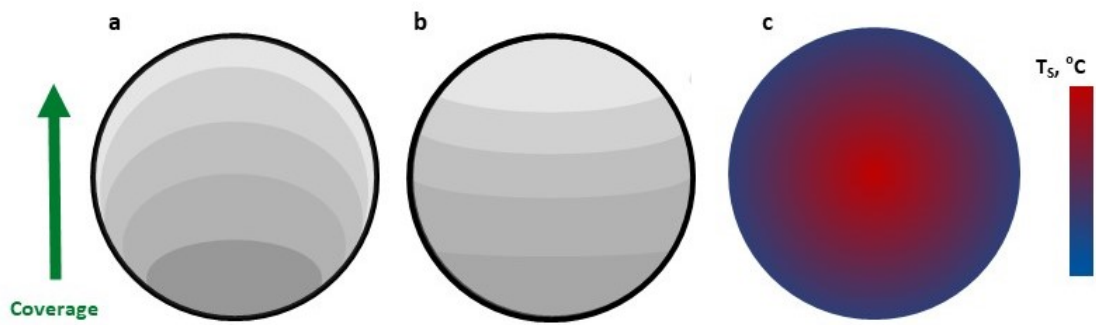

Fig. A1. Schematic of substrate coverage without (a) and with (b) a temperature gradient. (c) Schematic illustration of the substrate temperature gradient.

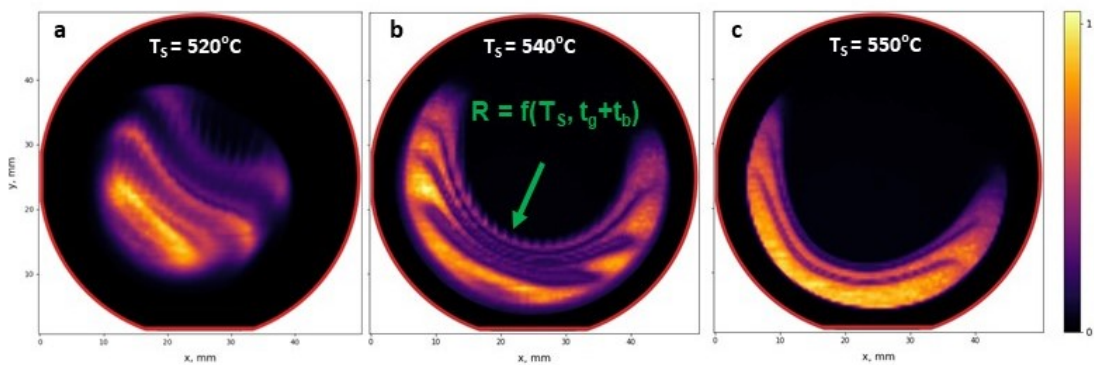

Fig. A2. False color integral PL maps of the InAs/InGaAs QD emission acquired at RT from the 2-inch GaAs(001) wafers. Rotation of samples and InAs sub-ML deposition were initiated at the same time. All samples shown were grown by aligning the In-cell towards -30° relative to the primary flat. Each cycle consisted of $t_g$ = 3 s followed by $t_p$ = 15 s break at 10 RPM. (a) Sample is grown by depositing 15 InAs at $T_S$ = 520°C, while (b) and (c) are grown by depositing 17 cycles at $T_S$ = 540°C and 550°C, respectively.

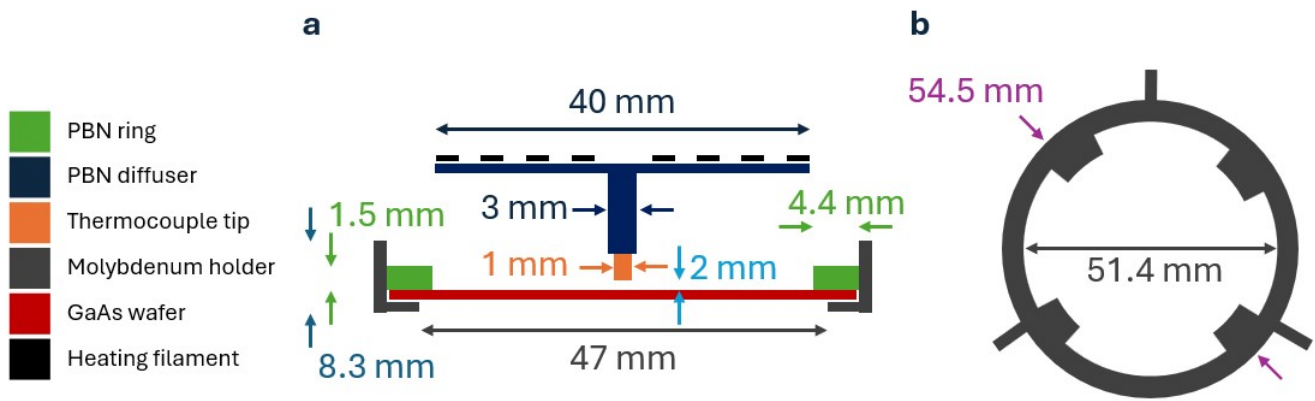

Fig. A3. a. Schematic cross-sectional view of a manipulator with a GaAs substrate placed on the holder ledge and fixed using the backing PBN ring in the employed MBE system. b. Schematic of the molybdenum holder.

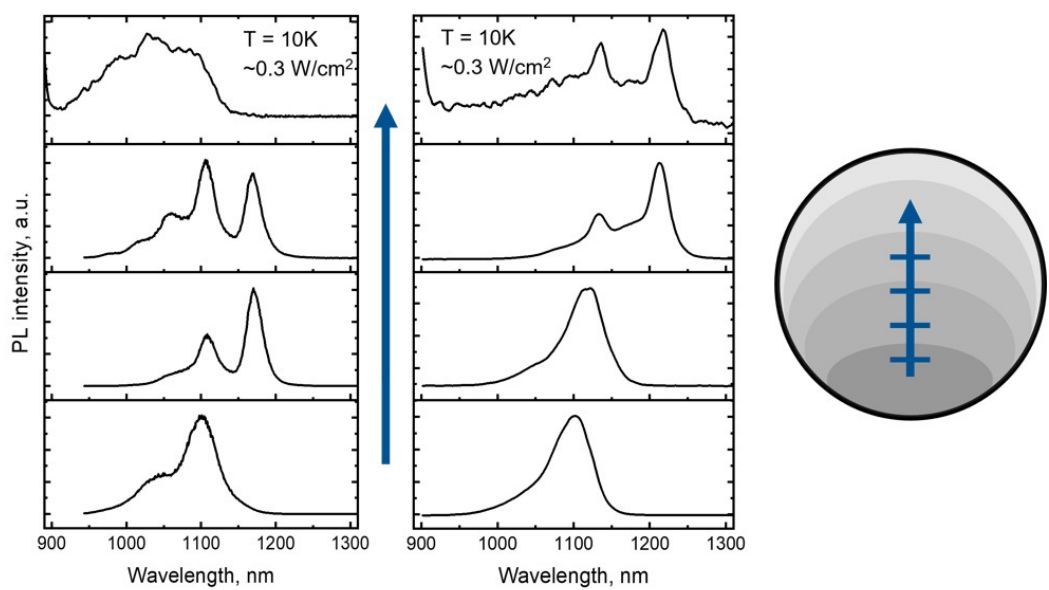

Fig. A4. PL spectra of InAs/GaAs QD acquired at T = 10 K and excitation power density of 0.3 W/cm² are shown along the InAs coverage gradient: (**a**) created using InAs sub-ML deposition without substrate rotation and (**b**) gradient deposition with synchronized substrate rotation and sub-ML deposition. Both samples were grown by depositing 12 InAs cycles at $T_S$ = 520°C and $P_{As}$(BEP) = $7 \times 10^{-6}$ Torr. Each cycle includes $t_g$ = 3 s followed by $t_p$ = 15 s. The arrow schematically shows the direction from high to low InAs coverage.

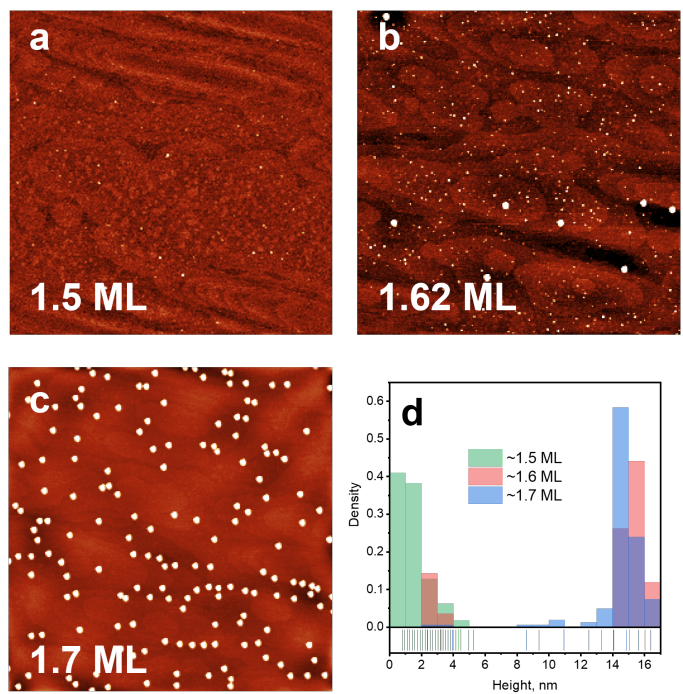

Fig. A5. 2×2 µm² AFM images for InAs coverage with GaAs PDL $\theta \sim 1.5$ ML (**a**), 1.62 ML (**b**) and 1.7 ML (**c**) along the gradient direction, respectively. **d.** Kernel density histogram represents estimation of the probability InAs QDs height density function for the given coverage regions: 1.5 ML (green), 1.6 ML (red) and 1.7 ML (blue)

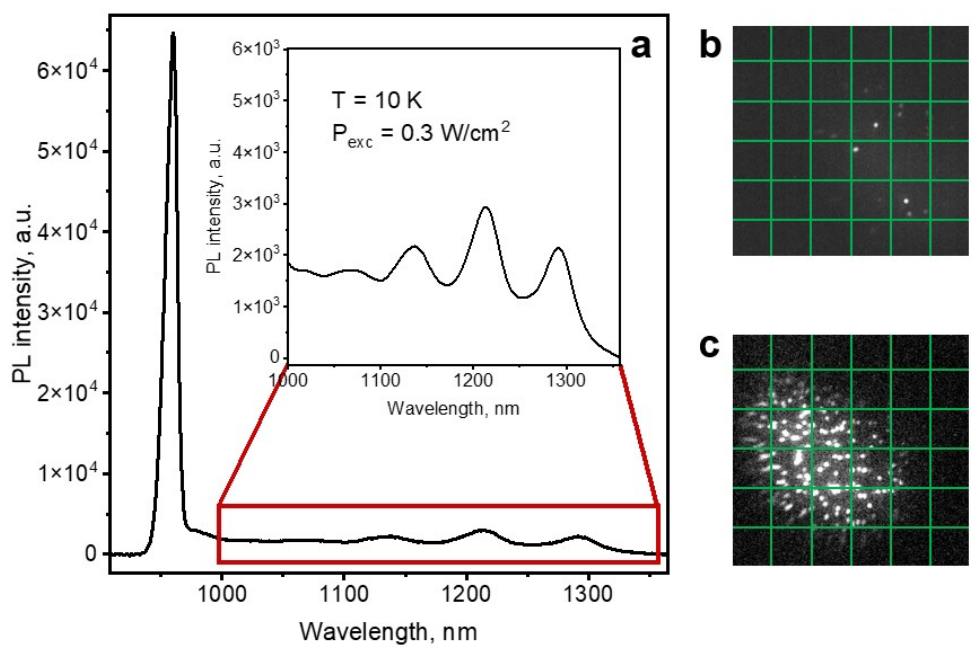

Fig. A6. **a.** Typical ensemble state-filling PL spectra of InAs/InGaAs QDs acquired in low-density regions at 10 K with an excitation power density of 0.3 W/cm². The inset shows the magnification of the region with InAs/InGaAs QDs emission. **b** and **c** Example of PL images from single InAs/InGaAs QDs along the density gradient measured using LED illumination. The spatial distance between the taken points where images are acquired is 1.5 mm. The image grid is 1×1 μm²

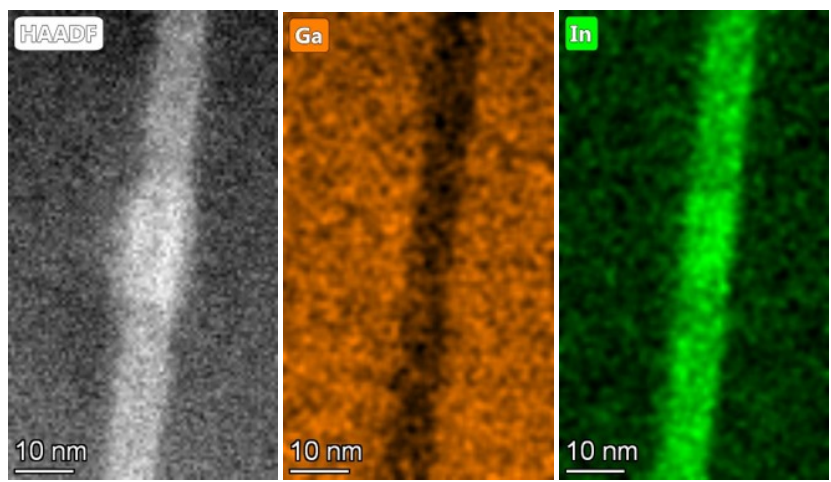

Fig. A7. High-Angle Annular Dark-Field (HAADF) and energy dispersive X-ray (EDX) mapping of elemental composition across the InAs/InGaAs QD layer, highlighting Ga and In distribution.

TABLE I. Voigt fit summary for multiple spectral regions and excitation powers.

| Fraction of $P_{sat}^{CX}$ | Excitonic state | FWHM (μeV) |
|---|---|---|
| 0.03 | CX | 27.0 |
| 0.03 | X | 22.3 |
| 0.33 | XX | 25.8 |
| 0.33 | CX | 26.2 |
| 0.33 | X | 23.4 |
| 1 | XX | 27.9 |
| 1 | CX | 27.2 |
| 1 | X | 24.2 |
| 1.3 | XX | 28.0 |
| 1.3 | CX | 27.4 |
| 1.3 | X | 24.3 |